# HEATING BY DISSIPATION OF SOUND WAVES IN THE INTERSTELLAR GAS

Miguel H. Ibañez Sanchez[1], Sandra M. Conde C[2]., and Pedro L. Contreras E. [2,3]
Address: [1]Centro de Física Fundamental, Universidad de Los Andes, Mérida, 5101 Venezuela, [2]Formally at Centro de Física Fundamental, Universidad de Los Andes, Mérida, 5101 Venezuela. [3]Departmento de Física, Universidad de Los Andes, Mérida, 5101 Venezuela,

Corresponding author email: pcontreras@ula.ve

**ABSTRACT**

The equilibrium resulting in a recombining plasma with arbitrary metallicity $Z$, heated by a mean radiation field $E$ as well as by sound waves dissipation due to thermal conduction, dynamic and bulk viscosity is analyzed. Generally, the heating by acoustic waves dissipation induces drastic changes in the range of temperature where the thermochemical equilibrium may exist. An additional equilibrium state appears which is characterized by a lower ionization and higher gas pressure than the equilibrium resulting when the wave dissipation is neglected. The above effects are sensibly to the values of the gas parameters as well as the wavelength and intensity of the acoustic waves. Implications in the interstellar gas, in particular, in the high velocity clouds are outlined.

## INTRODUCTION

It has been well established (Landau et al.,1987; Mihalas et al.,1984; Stix, 1992) that when the gas dynamic equations are linearized, assuming small disturbances, a rotational and a potential mode result which are independent each other. If viscosity is accounted for, it produces damping on the potential mode and if thermal conduction is taken into account an additional damped thermal mode appears (Landau et al.,1987).

The above two dissipative effects have been invoked as one of the mechanisms of heat input in different astrophysical plasmas, particularly, in the solar, and more generally in stellar atmospheric plasmas (Bird,1964; Narain et al.,1990; Stein et al,1972; Stein et al.,1974), in the interstellar medium (ISM) (Spitzer,1978; Spitzer,1982, Spitzer,1990), and more recently in the intracluster gas (Fabian et al.,2003; Fabian et al.,2005; Ruszkowski et al.,2004; Ferland et al.;2009; Ibañez et al.,2005; Fajardo et al.,2021).

Most of the works on this subject does not consider bulk viscosity dissipation, none the less the corresponding dissipation can be larger than both the dynamic viscosity and the thermal conduction dissipation, when chemical precesses are present (including ionization and recombination precesses) (Ibañez et al.,1993; Ibañez et al.,2019). In addition to the quantitative changes on the acoustic dissipation, the bulk viscosity also introduces important qualitative changes, in particular due to the fact that the bulk viscosity coefficient is dispersive (depends on the wave frequency), as it will be shown later on.

The present work is aimed to analyze how much a thermochemical chemical equilibrium is modified if heating by sound wave dissipation is taken into account in an homogeneous recombining plasma able to cool down by the well known cooling function for plasmas with arbitrary metallicity $Z$, ionized and heated by a mean radiation field $E$ as well as by dissipation of the acoustic waves. Implications in the structure the interstellar medium (ISM) and in high velocity clouds (HvCs) will be outlined.

Non-homogeneity effects will be neglected and only three dissipative mechanisms will be considered, i.e. the thermal conduction $\kappa$, the dynamic viscosity $\eta$, and the bulk viscosity $\zeta$. Molecular formation which appears once the hydrogen molecule $H_2$ forms (in gas phase or adbortion on cool grains), in particular the formation of the CO molecule (which is a much more stronger coolant than the corresponding atomic cooling), as well as the strong cooling by solid grains and their important opacity effects will not be taken into account at the present approximation. These additional effects completely change the radiative heating and cooling of the interstellar gas. These complications as well as magnetohydrodynamic effects will be carried out elsewhere (Fajardo et al.,2021).

**BASIC EQUATIONS**

For a fluid where a chemical reaction of the form $\sum_i b_i C_i = 0$ is proceeds, the equations of gas dynamic can be written in the form

$$\frac{\partial \rho}{\partial t} + \frac{\partial \rho v_i}{\partial x_i} = 0, \quad (1)$$

$$\rho \left( \frac{\partial v_i}{\partial t} + v_j \frac{\partial v_i}{\partial x_j} \right) = -\frac{\partial p}{\partial x_j} + \frac{\partial}{\partial x} \left[ \eta \left( \frac{\partial v_i}{\partial x_j} + \frac{\partial v_j}{\partial x_i} - \frac{2}{3} \delta_{ij} \frac{\partial v_k}{\partial x_k} \right) \right] + \frac{\partial}{\partial x} \left( \zeta \frac{\partial v_j}{\partial x_j} \right), \quad (2)$$

$$\frac{dT}{dt} + X(\rho, T, \xi) = 0, \quad (3)$$

$$A(\xi) \frac{dT}{dt} - \frac{p}{\rho^2} \frac{d\rho}{dt} + RTB(\xi, T) \frac{d\xi}{dt} + \mathcal{L}(\rho, T, \xi) - \frac{1}{\rho} \nabla \cdot (\kappa \nabla T) = 0, \quad (4)$$

$$p = \frac{R \rho T}{\mu(\xi)}, \quad (5)$$

where $d/dt$ is the convective derivative, $X(\rho, T, \xi)$ is the net rate function, $\mathcal{L}(\rho, T, \xi)$ the net rate of cooling per unit mass and time, which are functions of the degree of ionization $\xi$ (parameter determining the advance of the the reaction), the density $\rho$, and the temperature $T$. $\kappa$ is the thermal conduction coefficient, $\eta$ is the dynamic viscosity, and $\zeta$ the bulk viscosity (complex in this case). Furthermore an ideal gas state equation (5) has been assumed. $R$ is the universal gas constant, and the coefficients $A$, $B$ and $\mu$ are defined as follows

$$A(\xi) = \sum_i \frac{b_i \xi + x_i^0}{\gamma_i - 1}, \quad B(T) = \sum_i \left( \frac{b_i}{\gamma_i - 1} + \frac{1}{k_B T} b_i \epsilon_i^0 \right), \quad \frac{1}{\mu(\xi)} = \sum_i b_i \xi + x_i^0, \quad (6)$$

where $x_i^0$ is initial concentration, $\gamma_i$ and $\epsilon_i^0$ are the specific heat ration and zero point energy of the *i-esime* gas component, respectively, $k_B$ is the Boltzmann constant. Additionally, for a plasma with a degree of ionization $\xi$, the thermal conduction becomes

$$\kappa = 2.5 \times 10^3 (1 - \xi) T^{1/2} + 1.84 \times 10^{-5} \frac{\xi T^{5/2}}{\ln \Lambda_s(\rho, T, \xi)}, \quad (7)$$

where $\ln \Lambda_s(\rho, T, \xi)$ is the relation between the Debye screening, and the impact particles parameter (Braginskii, 1965; Spitzer 1962). The dynamic viscosity following (Braginskii, 1965; Spitzer 1962) takes the form

$$\eta = 2.21 \times 10^{-15} \frac{T^{5/2}}{\ln \Lambda_s(\rho, T, \xi)}. \quad (8)$$

On the other hand, the complex bulk viscosity $\zeta$ becomes

$$\zeta = \frac{\rho \tau}{1 - i\omega\tau} [c_\infty^2 - c_0^2], \quad (9)$$

where $\tau$ is the chemical relaxation time

$$\tau = \left( \frac{\partial X(\rho, T, \xi)}{\partial \xi} \right)^{-1}, \quad (10)$$

and the sound velocities $c_\infty^2$ and $c_0^2$ are defined as

$$c_\infty^2 = \left(\frac{\partial p}{\partial \rho}\right)_\xi \text{ and } c_0^2 = \left(\frac{\partial p}{\partial \rho}\right)_{eq} = \left(\frac{\partial p}{\partial \rho}\right)_\xi + \left(\frac{\partial p}{\partial \xi}\right)_\rho \left(\frac{\partial \xi_0}{\partial \rho}\right). \quad (11)$$

Here $\xi_0$ is the value of the chemical parameter at chemical equilibrium and $\left(\frac{\partial \xi_0}{\partial \rho}\right)$. In this work the c.g.s. unit system is used.

## LINEAR WAVES

If the velocity field of sound disturbances with wave number $k$ and frequency $\omega$ is assumed to be $v = v_x = v_1 \cos(kx - \omega t)$, and $v_y = v_z = 0$ as a first approximation, the time averaged in a volume $V$ of energy dissipation by the bulk viscosity $\zeta$ of sound waves becomes

$$\frac{1}{V} \int \overline{\zeta \frac{\partial v}{\partial x}} \, dV = \zeta_R \frac{v_1^2}{2} k^2, (12)$$

as can be realized easily, the real part $\zeta_R$ of the bulk viscosity $\zeta$ is

$$\zeta_R = \frac{\rho \tau}{1 + (\omega \tau)^2} [c_\infty^2 - c_0^2] \quad (13)$$

On the other hand, we have that

$$c_\infty^2 - c_0^2 = \left(\frac{\partial p}{\partial \xi}\right)_\rho, (14)$$

where $\left(\frac{\partial \xi}{\partial \rho}\right)_{eq}$ is calculated at equilibrium. This result will allow to calculate the bulk viscosity coefficients for reacting gases, as it will be shown later on.

Therefore, the heat input due to dynamic as well as bulk viscosity (Landau et al.,1987), and the thermal conduction becomes

$$\Gamma_\omega(\rho, T, \xi) = \gamma_d \frac{v_1^2}{2} k^2, (15)$$

where

$$\gamma_d = \left[\left(\frac{4}{3}\eta + \zeta_R\right) + \frac{(\gamma - 1)\kappa}{c_P}\right], (16)$$

and $c_P$ being the specific heat at constant pressure.

## THERMAL EQUILIBRIUM OF A PHOTOIONIZED GAS

Commonly the heat/loss function for a low density plasma at equilibrium temperatures (neglecting molecular formation as quoted out in the introduction) in the range $30 \, K < T < 3 \times 10^4 K$ can be written as

$$\rho \, \mathcal{L}(\rho, T, \xi) = \Lambda(\rho, T, \xi) - \Gamma_0(\rho, T, \xi) - \Gamma_\omega(\rho, T, \xi) \, (erg \, cm^{-3} \, s^{-1}), (17)$$

where $\Lambda(\rho, T, \xi)$ is the cooling rate and $\Gamma_0(\rho, T, \xi)$ is the heating input (different from wave dissipation) per unit volume and time, respectively. So, at thermal equilibrium, $\mathcal{L}(\rho, T, \xi) = 0$, i.e. the cooling rate becomes

$$\Lambda(\rho, T, \xi) = \Gamma_0(\rho, T, \xi) + \gamma_d \frac{v_1^2}{2} k^2. (18)$$

For an optically thin hydrogen plasma with metallicity $Z$ heated and ionized by a background radiation field of mean photon energy $E$, and an ionization rate $\varsigma$, the net rate function $X(\rho, T, \xi)$ and the cooling and heating rates rate per unit volume and time are respectively given by (Corbelli et al., 1995),

$$X(\rho, T, \xi) = N_0\, \rho\, [\xi^2 \alpha - (1 - \xi)\, \xi\, \gamma_c] - (1 - \xi)(1 + \phi)\, \varsigma, (19)$$

$$\Lambda(\rho, T, \xi) = (N_0\, \rho)^2 [\, (1 - \xi)\, Z\, \Lambda_{HZ} + \xi\, Z\, \Lambda_{eZ} - (1 - \xi)\, \Lambda_{eH} + \xi^2 \Lambda_{eH^+}\, ], (20)$$

and

$$\Gamma_0(\rho, T, \xi) = N_0\, \rho\, (1 - \xi)\, \varsigma\, [E_h + (1 + \phi)\, \chi_h], (21)$$

where $\phi$ denotes the number of secondary electrons, $E_h$ the heat released per photoionization (Shull et al., 1985), $\Lambda_{HZ}, \Lambda_{eZ}, \Lambda_{eH}$ and $\Lambda_{eH^+}$ respectively are the cooling efficiencies by collisions of neutral hydrogen-ions and metal atoms (Launay et. al., 1977; Dalgamo et al., 1972), electrons-ions and metal atoms (Dalgamo et al., 1972), $Ly\alpha$ emission by neutral hydrogen (Spitzer, 1978), and hydrogen recombination, on the spot approximation (Seaton, 1959).

From (Corbelli et al., 1995), follows that at ionization equilibrium state $X(\rho, T, \xi) = 0$, therefore

$$\left(\frac{\partial \xi}{\partial \rho}\right)_{eq} = -\frac{1}{2} \frac{(1 + \phi)\, \varsigma\, \left[\gamma_c + 2\alpha - \sqrt{F} + (1 + \phi)\, \varsigma / (N_0\, \rho)\right]}{(\alpha + \gamma_c)\, N_0\, \rho^2\, \sqrt{F}}, (22)$$

where

$$F = \gamma_c^2 + 2\, (\gamma_c + 2\alpha)\, \frac{(1 + \phi)\, \varsigma}{N_0\, \rho} + \left(\frac{(1 + \phi)\, \varsigma}{N_0\, \rho}\right)^2.$$

On the other hand, the damping scale length $l_d$ is given by

$$l_d = \frac{2\, \rho\, c_0}{\gamma_d\, k^2}, (23)$$

where equation (18) holds as far as the damping scale-length is larger than the sound wave length $\lambda = 2\pi/k$ i.e.

$$\frac{\lambda}{2\pi\, l_d} = \frac{\sqrt{\gamma_d\, \phi}}{\sqrt{2}\, \epsilon\, c_0}, (24)$$

where $= v_1/c_0$ (notice that $\epsilon$ does not have units); so, the dimensions of the region heated by sound waves dissipation is larger than the sound wavelength and the above approximation holds.

**APPLICATIONS**

**Interstellar gas (ISM)**

At the present subsection the results obtained previously will be applied to a gas with characteristic values of the parameters representative of the interstellar medium, i.e $N_0\, \rho = 1, Z = 1,$ and $E = 10^2 eV$, and in the range of temperature where the hydrogen ionization-recombination takes place.

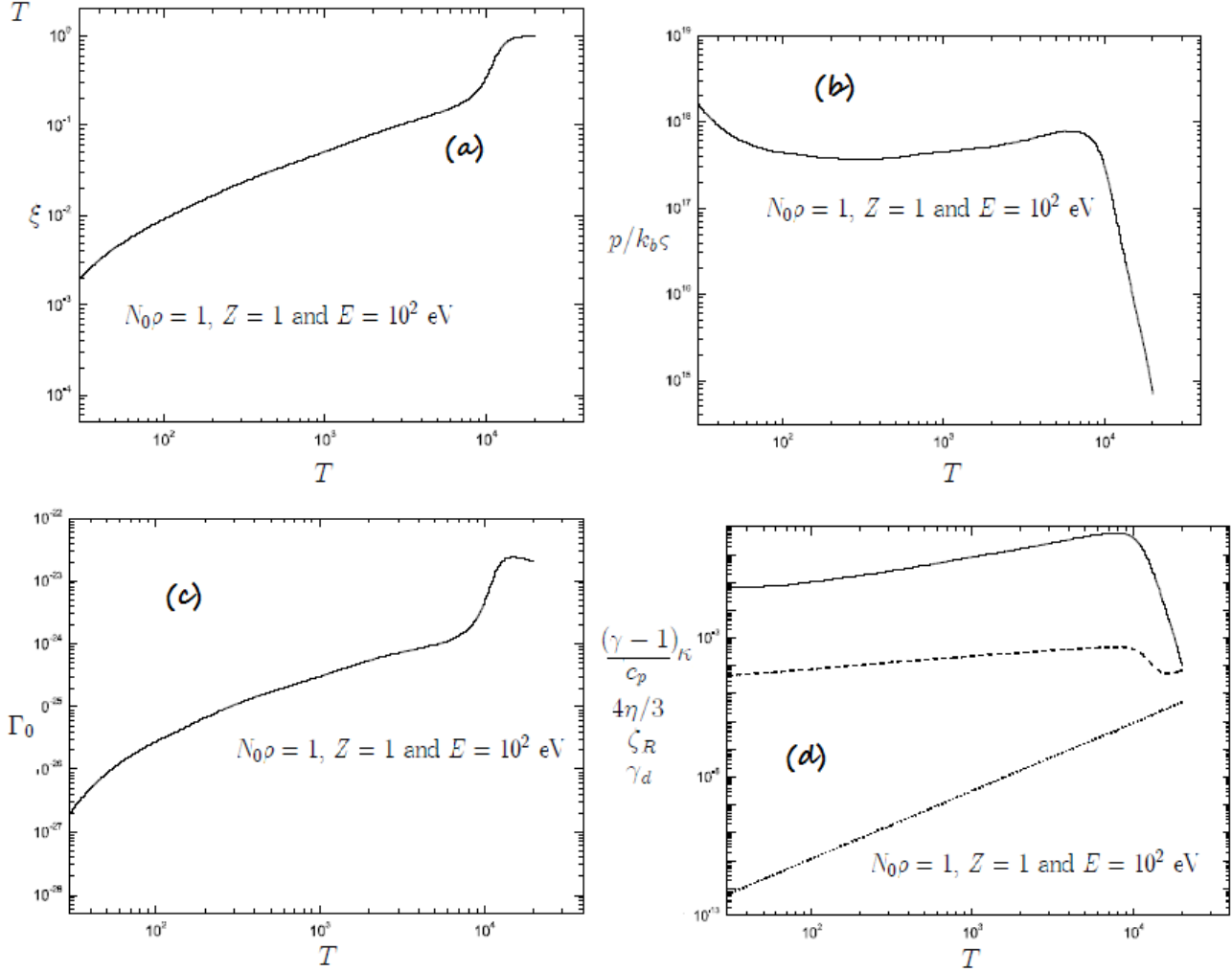

*Figure 1. The ionization ξ (a), the pressure at equilibrium (b), the heating input $\Gamma_0$ (c), and four dissipative terms (d) for a gas with $N_0 \rho = 1, Z = 1,$ and $E = 10^2 eV$ when $\epsilon k = 0\ cm^{-1}$ for a plasma at equilibrium i.e., $X(\rho,T,\xi) = 0,$ and $\mathcal{L}(\rho,T,\xi) = 0$. The dissipation $\gamma_d$ is undistinguished from the values of the real part of the bulk viscosity $\zeta_R$ at the scale of the Fig. 1(d).*

Figure 1 is a plot of the ionization $\xi$ (a), the pressure $p/(k_B \varsigma)$ (b), the heat input $\Gamma_0(c)$, and the dissipative coefficients $(\gamma - 1)\kappa/c_P$ (dash line in d), $4/3\ \eta$ (point line in d), $\zeta_R$ and $\gamma_d$ (thick line in d) without heat input by sound waves dissipation, ($\epsilon k = 0\ cm^{=1}$), and for a plasma at equilibrium i.e., $X(\rho,T,\xi) = 0,$ and $\mathcal{L}(\rho,T,\xi) = 0$.

For the above particular values of density, metallicity and mean photon energy, the thermal equilibrium may only exist for $T < 2.254 \times 10^4\ K$, the nature of the equilibrium for this particular case has been analyzed elsewhere approximation (Ibañez, 2009). As it can be seen in Fig. 1(d), the dynamic viscosity term (point line), and the thermal conduction term (dash line) are very small respect to the bulk viscosity $\zeta_R$ (thick line) which determines the total dissipation $\gamma_d$, that is undistinguished from $\zeta_R$ at the scale of Fig. 1(d).

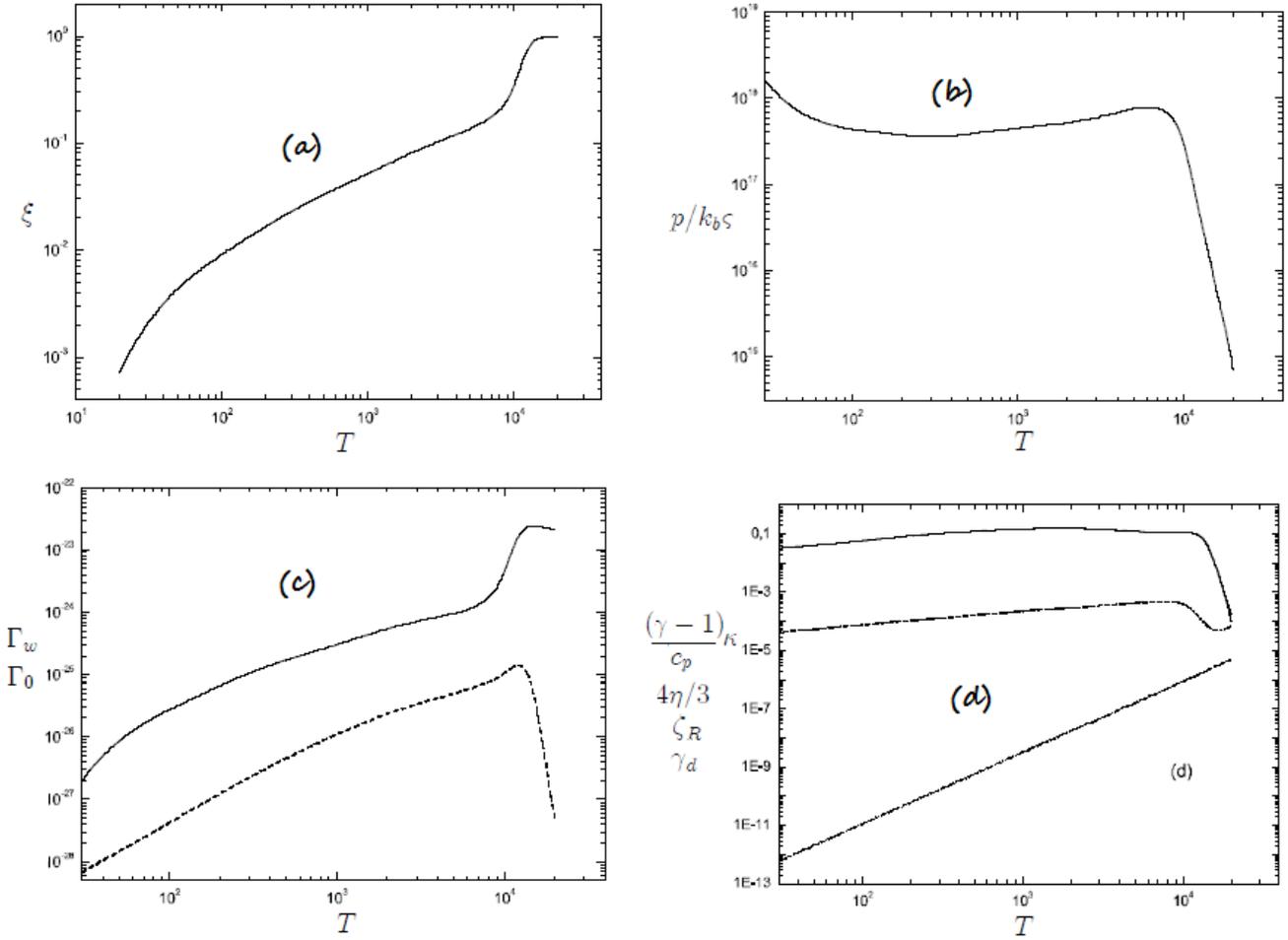

Figure 2. The ionization $\xi$ (a), the pressure (b), the radiative heating $\Gamma_w$ and the heating input $\Gamma_0$ (c), and four dissipative terms (d) for a gas with $N_0 \rho = 1, Z = 1, and\ E = 10^2 eV$ when $\epsilon k = 10^{-18} cm^{-1}$. The dissipation $\gamma_d$ again is undistinguished from the values of the real part of the bulk viscosity $\zeta_R$ at the scale of the Fig. 2(d).

When one considers sound waves dissipation the above results can drastically change depending on the value of the sound wavelength, more exactly on the value of $\epsilon k$. In fact, for the above values of parameters of the gas, sound waves with $\epsilon k \lesssim 10^{-18} cm^{-1}$ only produce small changes on variables the of interest, in particular at high temperatures $T > 10^3\ K$ as it is apparent in Figure 2, which are plots as those on Fig. 1 but for $\epsilon\ k = 10^{-18} cm^{-1}$. This is due to the fact that the heat input $\Gamma_0$ by dissipation of sounds shown by the dash line in Fig. 2(c), at this wavelengths, is less than one order of magnitude than the radiative heating $\Gamma_w$ seen in the thick line of Fig. 2(c).

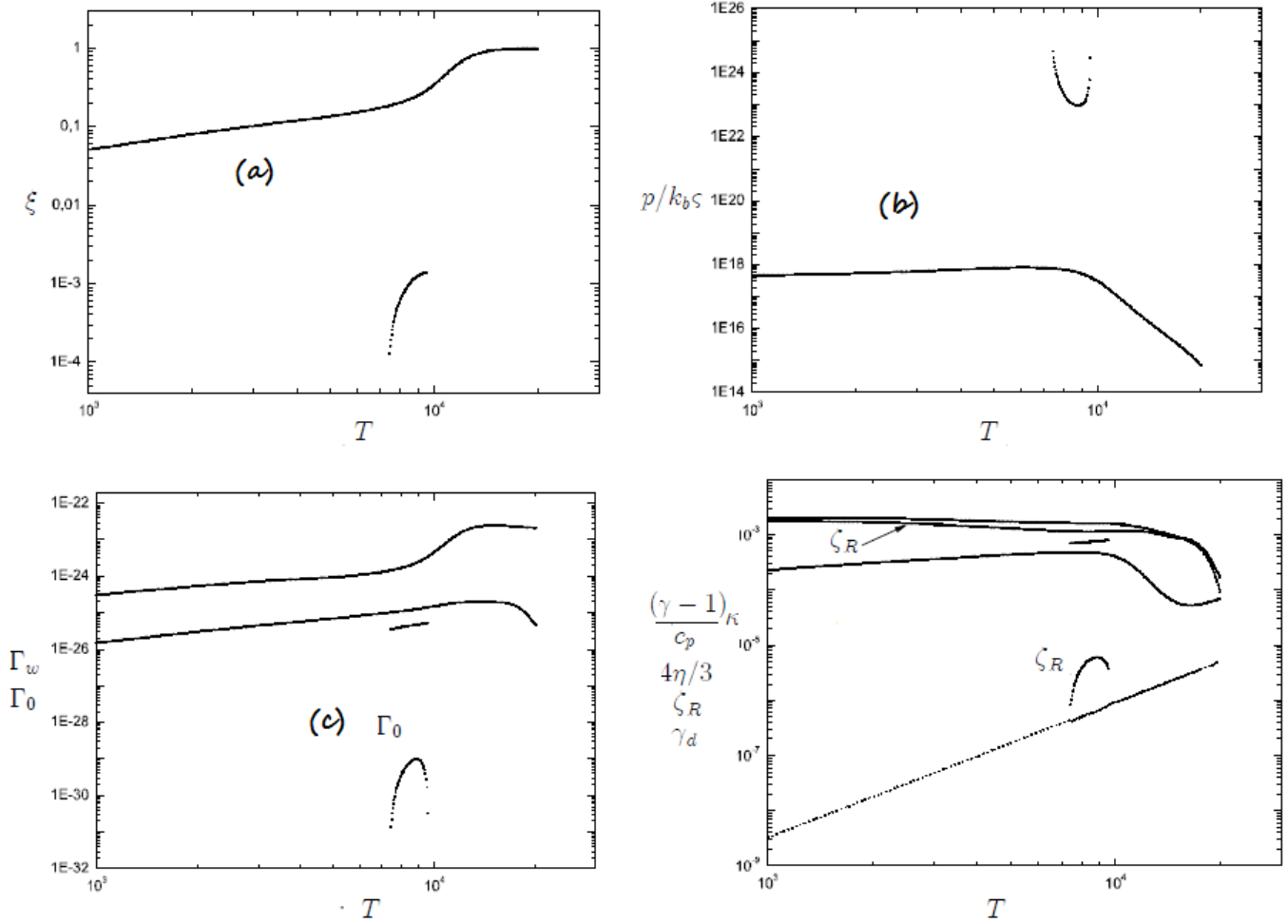

*Figure 3. The ionization $\xi$ (a), the pressure (b), the radiative heating $\Gamma_w$ and the heating input $\Gamma_0$ (c), the four dissipative terms (d) for a gas when $\epsilon k = 10^{-17} cm^{-1}$.*

Figure 3 is similar to Fig. 2 but for $\epsilon k = 10^{-17} cm^{-1}$. At this shorter value of the acoustic wavelength, qualitative changes appear. In addition to a modified equilibrium ionization given by the upper branch in Fig. 3(a) by comparing with the case when wave heat input is neglected. Also, an additional ionization branch appears (lower branch in Fig. 3(a)) but this ionization corresponds to a higher pressure (upper branch in Fig. 3(b)) which occurs in a narrow interval of temperature $7 \times 10^3 < T < 10^4 \, K$.

These two values of ionization and pressure produce two values of the heating rates $\Gamma_0(\rho, T, \xi)$ and $\Gamma_\omega(\rho, T, \xi)$ Fig. 3(c) as well as two value in the dissipation coefficients $\kappa, \eta, and \, \zeta_R$ in Fig. 3(d) in the above interval of temperature. From the physical point of view, the above results imply that when the heat input by acoustic waves in a recombining gas becomes important, the gas can be at equilibrium in two different phases: one at high ionization and low pressure, and the another one at low ionization and high pressure for the same value of the density.

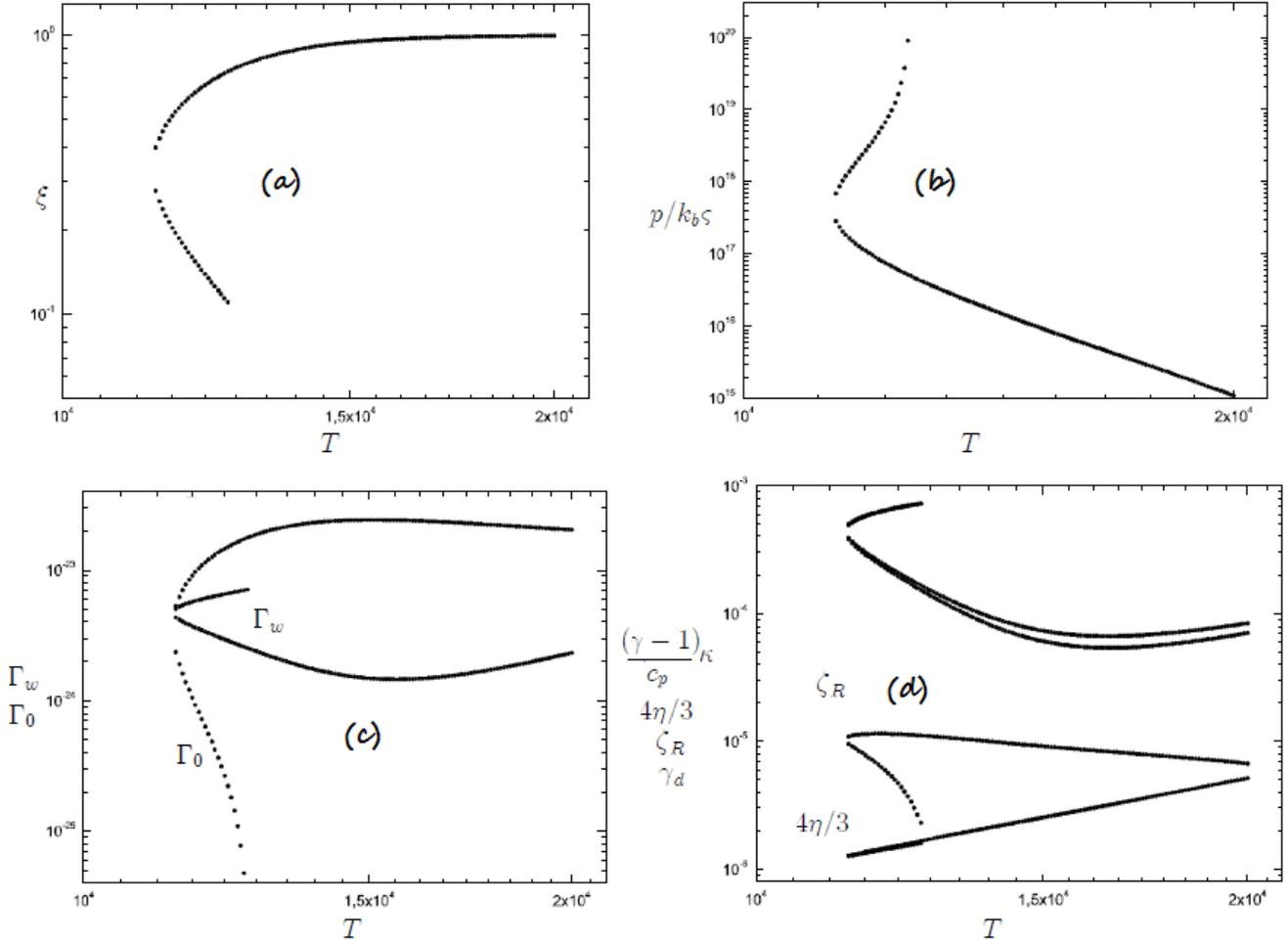

Figure 4. The ionization $\xi$ (a), the pressure (b), the radiative heating $\Gamma_w$ and the heating input $\Gamma_0$ (c), the four dissipative terms (d) for a gas when $\epsilon k = 10^{-16} cm^{-1}$. A birfurcation point is seen in all 4 figs. for $T = 1.14 \times 10^4$ K.

The effects shown in Fig. 3 shift towards higher values of the temperature when $\epsilon k$ increases as can be seen in Figure 4 which is as Fig. 3 but for $\epsilon k = 10^{-16} cm^{-1}$. For this particular value of $\epsilon k$, the ionization Fig. 4(a), pressure Fig. 4(b), both heating rates Fig. 4(c), as well as, the dissipative terms Fig. 4(d) show a bifurcation point at a temperature $T = 1.14 \times 10^4$ K.

A low ionization branch and high pressure branch appear in the range of temperature $1.14 \times 10^4 < T < 1.268 \times 10^4$ K. For these equilibrium states the heating $\Gamma_0(\rho, T, \xi) < \Gamma_\omega(\rho, T, \xi)$. Contrary to the high ionization (and pressure) branch were the radiation heating is higher than the wave heating in Figs. 4(a), 4(b), and 4(c). For this case, the thermal conduction dissipation terms dominates over the viscosity terms. The low ionization branch corresponds to a total dissipative branch $\gamma_d$ higher than that corresponding to higher ionization shown in Fig. 4(d).

When the value of $\epsilon k$ increases the interval of temperature where this equilibrium may exist decreases in such a way that the equilibrium can not exist for $\epsilon k > 5 \times 10^{-16} cm^{-1}$. Figure 5 shows the plot of the the variables under discussion have been plotted $\epsilon k = 5 \times 10^{-16} cm^{-1}$. For this particular value, the ionization and pressure becomes one-valued function of temperature (as in the case $\epsilon k = 10^{-18} cm^{-1}$) but the gas pressure shows a minimum value at $T = 1.849 \times 10^4$ K seen in Fig. 5(b). At this limiting value the acoustic dissipation $\Gamma_\omega(\rho, T, \xi)$ becomes higher than the radiation heating $\Gamma_0(\rho, T, \xi)$ as it is apparent from Fig. 5(c). Note that in this case, the dissipation is due mainly to thermal conduction, as for the previous $\epsilon k$ value, please see Fig. 5(d).

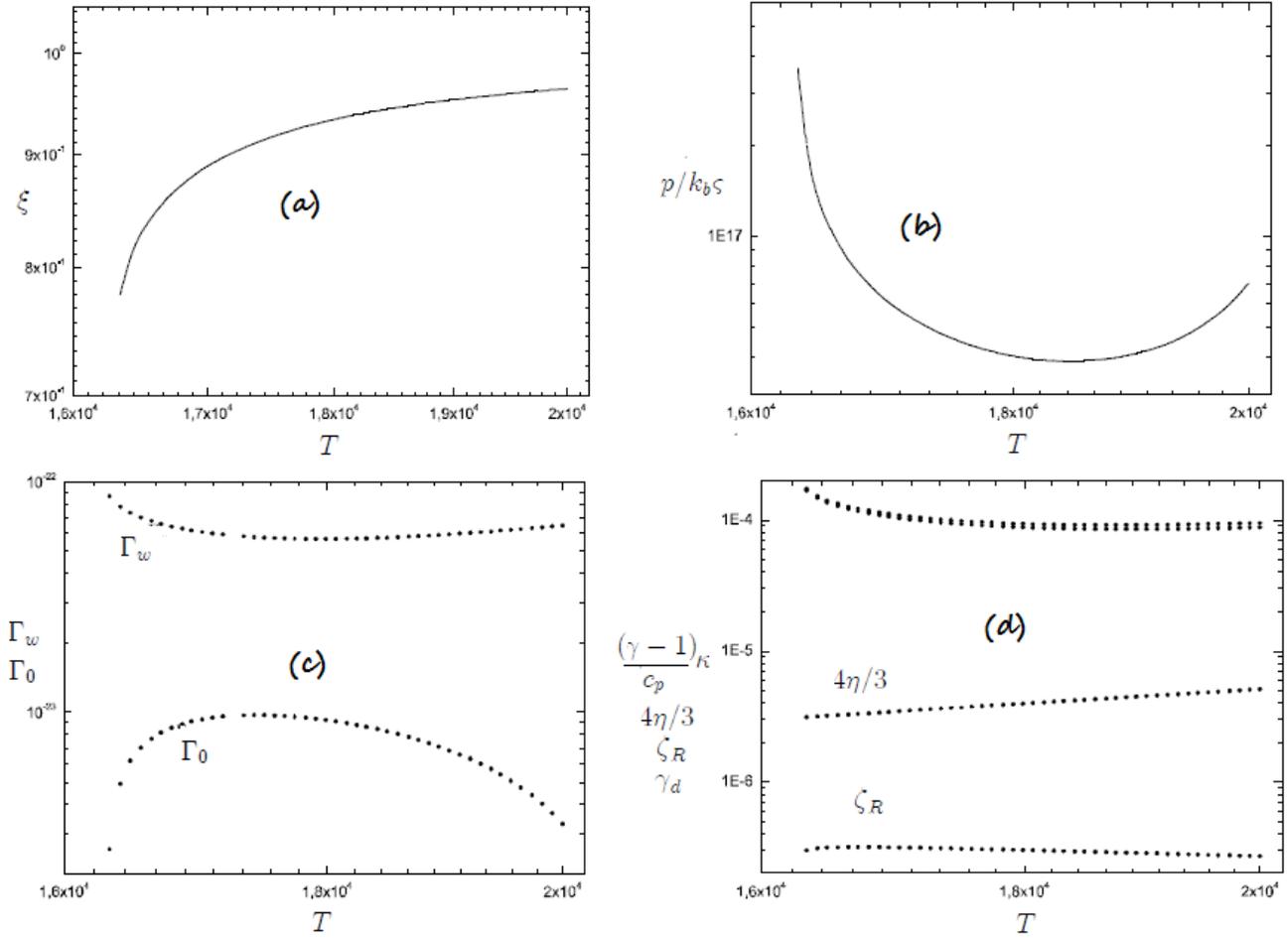

*Figure 5. The ionization $\xi$ (a), the pressure (b), the radiative heating $\Gamma_w$ and the heating input $\Gamma_0$ (c), the four dissipative terms (d) for a gas when $\epsilon k = 5 \times 10^{-16} cm^{-1}$. The higher plot in fig 5(d) correspons to thermal conduction.*

Figure 6 with the damping length scale $l_d$ is obtained for the following four values of the magnitude $\epsilon k$, i.e., $\epsilon k = 10^{-18} cm^{-1}$ (a), $\epsilon k = 10^{-17} cm^{-1}$ (b), $\epsilon k = 10^{-16} cm^{-1}$ (c), and $\epsilon k = 5 \times 10^{-16} cm^{-1}$ (d). The temperature dependence of the damping length scale $l_d$ (Eq. 24) is shown in the Figs 6(a), 6(b), 6(c) and 6(d), respectively.

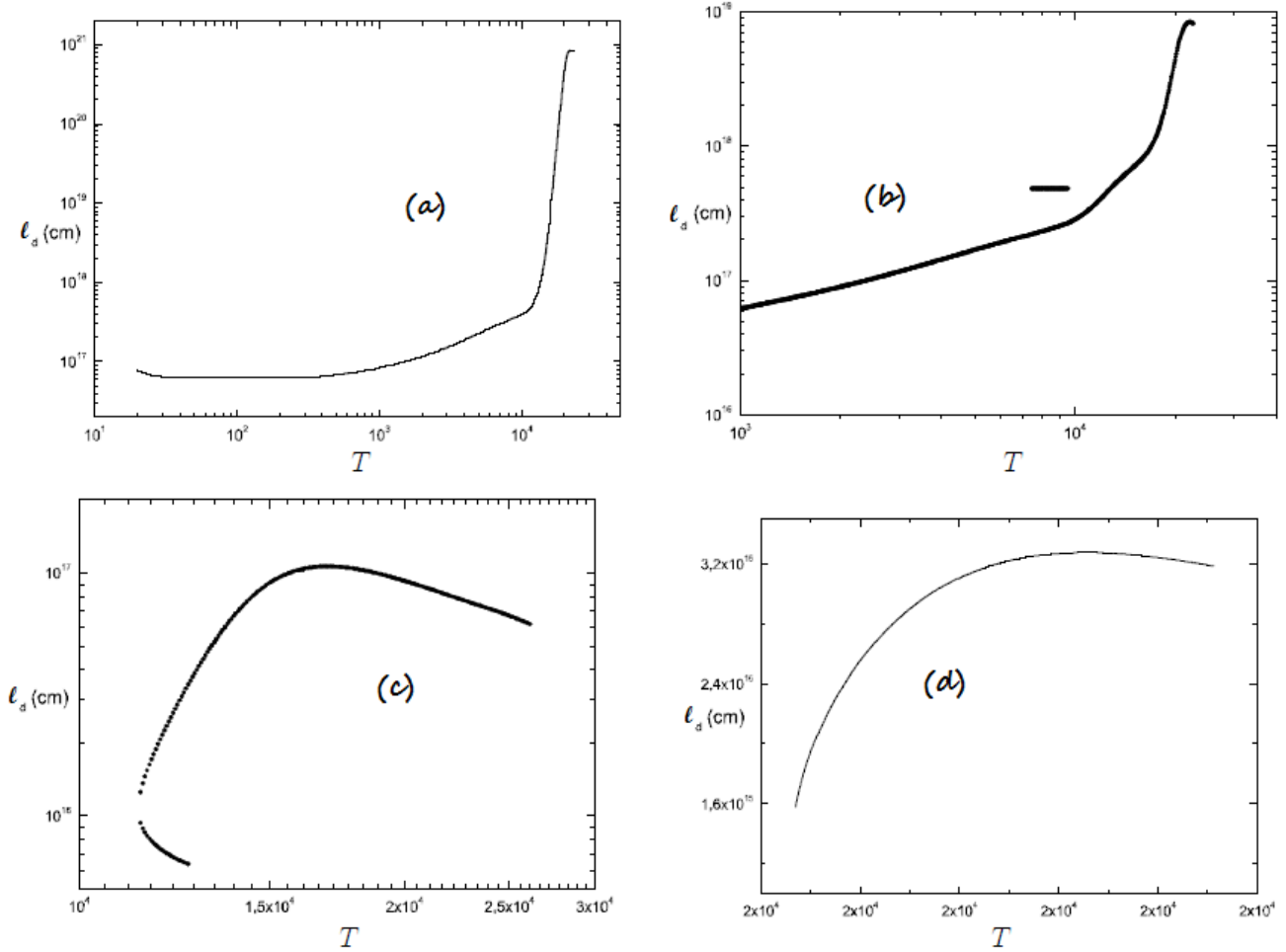

*Figure 6. The damping length scale $l_d$ as function of the temperature T, for the following four values of the magnitude $\epsilon k$, i.e., $\epsilon k = 10^{-18} cm^{-1}$ (a), $\epsilon k = 10^{-17} cm^{-1}$ (b), $\epsilon k = 10^{-16} cm^{-1}$ (c), and $\epsilon k = 5 \times 10^{-16} cm^{-1}$ (d).*

Figure 7 shows the plots for the corresponding ratios $\lambda/(2\pi l_d)$. For large wavelengths when the sound wave heat input is much more smaller than the radiative heating seen in fig 6(a), regions of the order of parsecs up to kiloparsecs can exist in thermochemical equilibrium in a wide range of temperature. When the wave number decreases this range of temperature decreases as well as the dimensions of the regions where such an equilibrium may exist, and the two equilibria become apparent.

Regions of high pressure are larger than those at lower pressure for $\epsilon k = 10^{-17} cm^{-1}$ Fig. 6(b) but the opposite occurs for $\epsilon k = 10^{-16} cm^{-1}$ in Fig. 6(c). Close to the limiting value were the equilibrium may exist only small regions, of the order of $10^{-3} pc$ and in a narrow interval of temperature can exist. As it is apparent from Figure 7 the present approximation holds, $\lambda/l_d < 1$ in the range of temperature where the recombination-ionization becomes important.

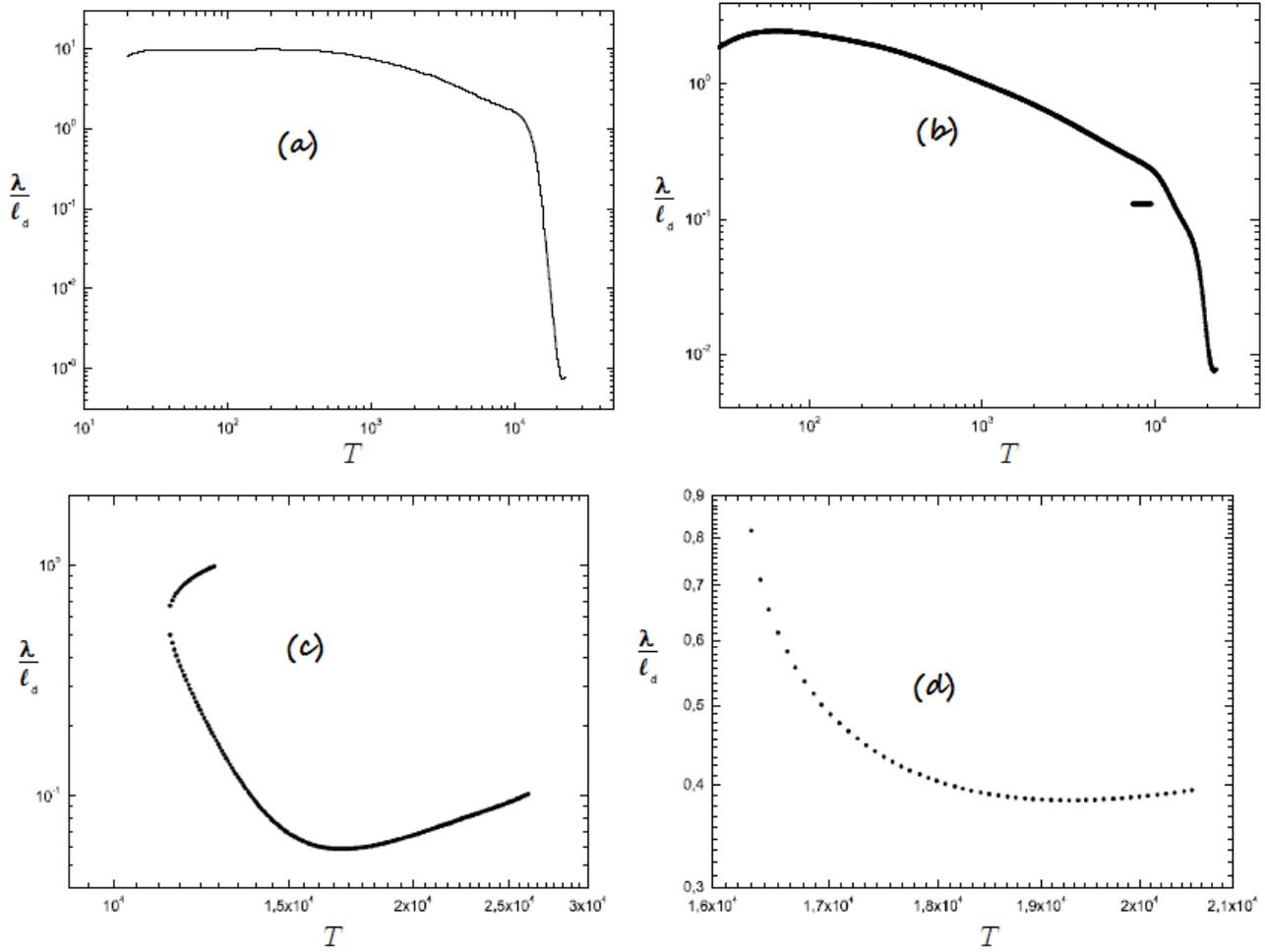

Figure 7. The ratios $\lambda/l_d$ for $\epsilon k = 10^{-18} cm^{-1}$ (a), $\epsilon k = 10^{-17} cm^{-1}$ (b), $\epsilon k = 10^{-16} cm^{-1}$ (c), and $\epsilon k = 5 \times 10^{-16} cm^{-1}$ (d).

**High velocity clouds at high galactic latitude (HvCs)**

Observations suggest that for the hydrogen gas at large galactic latitude, and in the halos of other galaxies the metallicity ranges between $0.03 \lesssim Z \lesssim 0.3$ $0.03 and a the mean photon energy between $0.3 \lesssim E \lesssim 2\ keV$ (Collins et al.,2004; Maller et al.,2004; Miller et al.,2015; Wakker et al., 2001; Wakker, 2004; Tripp et al.,2003; Olano, 2008; Lockman et al., 2008, Ibañez et al.,2011, Conde, 2010)

Thus equilibrium states may exist for $\epsilon k$ smaller than a threshold value $(\epsilon k)_{thr}$ which depends on the exact values of $Z$ and $E$ (see Figure 8).

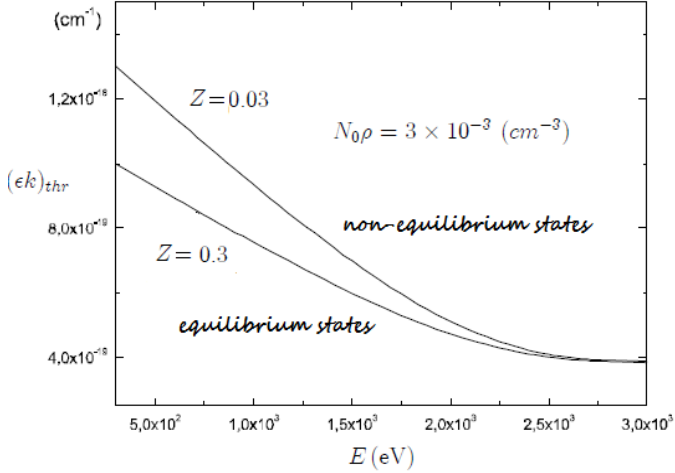

Figure 8. The threshold value $(\epsilon k)_{thr}$ as a function of the mean radiation field $E$ for the metallicity values $Z = 0.03, 0.3$ in HvCs with $N_0\, \rho = 3 \times 10^{-3}\, cm^{-3}$.

For $\epsilon k > (\epsilon k)_{thr}$ equilibrium states can not exit. Therefore, the hydrogen gas into High velocity Clouds (HvCs) can be in thermochemical equilibrium as far as the HvCs have dimensions $l > l_{thr}$, where $l_{thr} = 3.24 \times 10^{-19}\, (\epsilon k)_{thr}\, pc$ the above range of values of metallicity and mean photon energy. Strictly speaking this threshold value has to be taken as a limiting value, for real situations one should expect thermochemical equilibrium for gas clouds with dimensions $l \gg l_{thr}$. This conclusion can also be applied to the intracluster clouds, however this important case will be analyzed elsewhere.

Additionally, depending on the particular values of $Z$ and $E$, temperature gaps appear where the equilibrium can not exist. Say for instance, for $Z = 0.3$ and $E = 0.3\, keV$ the gap is between temperature $2.405 \times 10^4 < T < 8.149 \times 10^5\, K$, and for $Z = 0.03$ between $2.297 \times 10^4 < T < 2.255 \times 10^5\, K$, for this particular value of $Z$ only one equilibrium exists just at $T = 2.255 \times 10^5\, K$.

From the above results one may conclude that HvCs at thermal equilibrium may only exist in well defined of ranges of temperature, (depending on $E, Z$ and $N_0\rho$) and with dimensions $l \gg l_{thr}$.
.

**SUMMARY**

Generally, the dissipation of acoustic waves by thermal conduction, dynamic viscosity. as well as, bulk viscosity can introduce drastic quantitative and qualitative changes in the equilibrium of the interstellar gas. Depending on the particular values of the metallicity, the mean radiation energy input, and the particle density:

- The bulk viscosity can be the most important dissipation mechanism.

- Several equilibrium branches may appear in certain temperature ranges which depend on the particular values of $\epsilon k$.

- For large enough values of $\epsilon k$ the range of temperature where equilibrium may occurs becomes very narrow.

- For $\epsilon k$ values larger than a threshold value $(\epsilon k)_{thr}$ the equilibrium does not exist which imposes a limiting vale to the scale length of equilibrium structures, in particular, to the dimensions of HvCs.

Strictly speaking, in any plasma, in particular, in a very complex interstellar medium (ISM) a (discrete as well as continuos, depending on the origin of the waves) spectral distribution for sound waves should be present. However, determining for which particular $\epsilon k$ values of such spectrum the equilibrium ($X(\rho, T, \xi) = 0\, and\, \mathcal{L}\, (\rho, T, \xi)$=0) is modified by the waves dissipation is an important issue.

We emphasize that this work is essentially restricted to a linear approximation, in which the wave dissipation is a second order phenomena (Landau et al.,1987), for this, the results depend on $\epsilon k$ which allows one to get a "first" information on the dependence on the wave amplitude.

On the other hand, turbulence is by itself a non-linear phenomena, a problem out the scope of the paper. Certainly, turbulence should be present in the different "regions" of the ISM, which is a typical (very complex) plasma where many relaxation time scales go into play. Models beyond the qualitative ones (Kolmogorov-like) (Braun et al.,2012) do not exist, and the understanding of turbulent plasmas under realistic physical conditions is an open question.

**Authorship contribution statement**.

M. Ibañez: Conceptualization, Methodology, Software, Investigation, Validation, Writing – original draft, Supervision.
S. Conde: Methodology, Validation , Investigation, Writing – original draft.
P. Contreras: Investigation, Data curation, Visualization, Writing – Review & Editing

**Declaration of competing interest**

The authors declare that he has no known competing financial interests or personal relationships that could have appeared to influence the work reported in this paper.

**REFERENCES**

Bird, G. A. 1964. A gas-dynamic model of the outer solar atmosphere. The Astrophysical Journal. 139(2):684-689. DOI: http://dx.doi.org/10.1086/147794

Braginskii, SI. 1965. Transport processes in a plasma. Reviews of Plasma Physics. 1:205-217.

30. Braun H. and Schmidt W., 2012. A semi-analytic model of the turbulent multi-phase interstellar medium, Mon. Not. R. Astron. Soc. 42:1838, https://doi.org/10.1111/j.1365-2966.2011.19889.x

Corbelli, E., and Ferrara, A. 1995. Instabilities in photoionized interstellar gas. The Astrophysical Journal. 447(2):708-720.

Collins, J. A.; Shull, J. M.; and Giroux M. L. 2004. Highly Ionized High-Velocity Clouds toward PKS 2155–304 and Markarian 509. The Astrophysical Journal 605(1):216 DOI: 10.1086/382269

Conde C. Sandra M. 2010. Calentamiento de plasmas por ondas hidrodinamicas. M. Sc. Thesis, University of Los Andes, Center for Fundamental Physics.

Dalgarno A. and .McCray R. A. 1972. Heating and ionozation of HI regions, Ann. Rev. Astron. Astroph., 10, 375 -426.

Fabian, AC., Sanders, JS., Allen, SW., Crawford, CS., Iwasawa, K., Johnstone, RM., Schmidt, RW. and Taylor, GB. 2003. A deep Chandra observation of the Perseus cluster: shocks and ripples. Monthly Notices of the Royal Astronomical Society. 344(3):L43-L47. DOI: https://doi.org/10.1046/j.1365-8711.2003.06902.x

Fabian, AC., Reynolds, CS., Taylor, GB. and Dunn, RJH. 2005. Monthly Notices of the Royal Astronomical Society. 363(3):891-896. DOI: https://doi.org/10.1111/j.1365-2966.2005.09484.x

Fajardo M. Jorge J.; Ibañez S. Miguel H, and Contreras E. Pedro L. 2021. An eigenvalue analysis of damping in optical thin plasmas.Rev. Mex. Fis. 67 061502 1–13 DOI: https://doi.org/10.31349/RevMexFis.67.061502

Ferland, GJ., Fabian, AC., Hatch, NA., Johnstone, RM., Porter, RL., Van Hoof, PAM. and Williams, RJR. 2009. Collisional heating as the origin of filament emission in galaxy clusters. Monthly Notices of the Royal Astronomical Society. 392(4):1475-1502. DOI: https://doi.org/10.1111/j.1365-2966.2008.14153.x

Ibáñez, SMH. 2009. The stability of optically thin reacting plasmas: Effects of the bulk viscosity. The Astrophysical Journal. 695(1):479-487. DOI: https://doi.org/10.1088/0004-637X/695/1/479

Ibañez, S. M. H., and Escalona, O. B. 1993. Propagation of hydrodynamic waves in optically thin plasmas. The Astrophysical Journal, 415:335–341. DOI: https://doi.org/10.1086/173167

Ibañez,S. M. H., Shchekinov, Y., and Bessega, M. C. 2005. On the stability of subsonic thermal fronts. Phys. Plasmas, 12:1–7 DOI: https://doi.org/10.1063/1.1993287

Ibañez S. Miguel H., and Conde C. Sandra M.. 2011 Heating of a Recombining Plasma by Bulk Viscosity Dissipation of Sound Waves, Rev. Mex. Astron. Astrofis. Ser. Conf. 40:188

Ibañez S. Miguel H., and Contreras E. Pedro L. 2019. Damping of Sound Waves by Bulk Viscosity in Reacting Gases, Can. J. Pure Appl. Sci. 13(3) 4897-4906  ISSN: 1920-3853

Landau, L. D. and Lifshitz, E. M. 1987. Fluid Mechanics. Pergamon Press, London.

Launay J. M., and Roueff, E. 1977. Fine structure excitation of carbon and oxygen by atomic hydrogen impact. Astronomy and Astrophysics, 56(1-2) :289-292

Lockman F. J.; R. A. Benjamin R. A.; Heroux, A. J.; and Langston. G.I. 2008. The Smith Cloud: A High-Velocity Cloud Colliding with the Milky Way. The Astrophysicial Journal, 679(1), 21L DOI: 10.1086/588838

Maller A. H. and Bullock J. S. 2004. Multiphase galaxy formation: high-velocity clouds and the missing baryon problem. Monthly Notices of the Royal Astronomical Society, 355:694-712.

Mihalas D., and Mihalas B. W., 1984. Foundations of Radiation Hydrodynamics, Oxford University Press.

Miller, E. D.; and Bregman; B. P. 2015. Constarining the milky ways hot gas halo with 0 vii emission lines. The Astrophysical Journal, 800, 14. DOI: 10.1088/0004-637X/800/1/14

Narain, U. and Ulmschneider, P. 1990. Chromospheric and coronal heating mechanisms. Space Science Reviews. 54(3-4):377-445. http://dx.doi.org/10.1007/BF00177801

Olano, C. A. 2008. Distribution of the high-velocity clouds in the Galactic halo. Astronomy and Astrophysics, 485(2), 457-473. DOI https://doi.org/10.1051/0004-6361:20077556

Ruszkowski, M., Brüggen M. and Begelman, MC. 2004. Cluster heating by viscous dissipation of sound waves. The Astrophysical Journal. 611(1):158-163. DOI: https://doi.org/10.1086/422158

Seaton, MJ. 1959. Radiative recombination of hydrogenic ions. Monthly Notices of the Royal Astronomical Society. 119(2):81-89. DOI: https://doi.org/10.1093/mnras/119.2.81

Stein, RF., and Schwartz, RA. 1972. Waves in the solar atmosphere. II. Large-amplitude acoustic pulse propagation. The Astrophysical Journal. 177:807-828. DOI: http://dx.doi.org/10.1086/151757

Stein, RF., and Leibacher, J. 1974. Waves in the solar atmosphere. Annual Review of Astronomy and Astrophysics. 12:407-435. DOI: http://dx.doi.org/10.1146/annurev.aa.12.090174.002203

Shull, JM., and Van Steenberg, ME. 1985. X-ray secondary heating and ionization in quasar emission-line clouds. The Astrophysical Journal. 298(1, Part 1):286-274. DOI: http://dx.doi.org/10.1086/163605

Spitzer, L. 1978. Physical Processes in the Interstellar Medium. Interscience, John Wiley and Sons Inc., New York. pp333. DOI: http://dx.doi.org/10.1002/9783527617722

Spitzer, L. 1982. Acoustic waves in supernova remnants. The Astrophysical Journal. 262(1, Part 1):315-321. DOI: http://dx.doi.org/10.1086/160423

Spitzer, L. 1990. Theories of the hot interstellar gas. Annual Review of Astronomy and Astrophysics. 28:71-101. DOI: https://doi.org/10.1146/annurev.aa.28.090190.000443

Spitzer, L. 1962. Physics of Fully Ionized Gases. Interscience, John Wiley and Sons Inc., New York. pp61.

Stix, T. H. 1992. Waves in plasmas. Springer, Berlin.

Tripp, T.M., Wakker, B.P., Jenkins, E.B., Bowers, C.W., Danks, A.C., Green, R.F., Heap, S.R., Joseph, C.L., Kaiser, M.E., Linsky, J.L., and Woodgate, B.E. 2003. Complex C: A Low-Metallicity, High-Velocity Cloud Plunging into the Milky Way. The Astronomical Journal, 125:3122-3144

Wakker, B. P. 2001. Distamces and metallicities of high- and intermediate- velocity clouds Astrophysical J. Supplement, 136:463-535

Wakker, B. P. 2004. High velocity clouds and the local group. IAU Symposium Series. 217:2